\documentstyle[12pt]{article}

\newcommand{\be}{\begin{equation}}
\newcommand{\nn}{\nonumber}
\newcommand{\bea}{\begin{eqnarray}}
\newcommand{\eea}{\end{eqnarray}}
\newcommand{\ba}{\begin{array}}
\newcommand{\ea}{\end{array}}
\newcommand{\ee}{\end{equation}}

\expandafter\ifx\csname mathbbm\endcsname\relax

\else

\fi \textheight 22cm \textwidth 15cm \topmargin 1mm \oddsidemargin
5mm \evensidemargin 5mm

\def\a{\alpha}

\newcommand{\beas}{\begin{eqnarray*}}
\newcommand{\eeas}{\end{eqnarray*}}
\newcommand{\bes}{\begin{equation*}}
\newcommand{\ees}{\end{equation*}}

\newcommand{\lf}{\left}
\newcommand{\ri}{\right}
\newcommand{\f}{\frac}

\def\tr           {\mbox{\rm tr}\,}

\def\i2           {\mbox{$\frac{i}{2}$}}

\def\al           {\alpha}

\def\bet           {\beta}

\def\Mt           {\widetilde M}
\def\St           {\widetilde S}
\def\del           {\delta}

\def\ga           {\gamma}

\def\la           {\lambda}

\def\ph           {\phi}

\def\ps           {\psi}

\def\si           {\sigma}

\def\pl           {\partial}

\def\ran          {\rangle}
\def\lan          {\langle}

\begin{document}

\input epsf
\begin{titlepage}
\hfill \vbox{
    \halign{#\hfil         \cr
           hep-th/0302083 \cr
           IPM/P-2003/007 \cr
           } 
      }  
\vspace*{20mm}
\begin{center}
{\Large {\bf  ${\cal N}=2$ $SO(N)$ SYM theory from Matrix Model }\\
}

\vspace*{15mm} \vspace*{1mm} {Reza Abbaspur$^a$, Ali Imaanpur$^{a,b}$
and  Shahrokh Parvizi$^a$}

\vspace*{1cm}

{\it $^a$ Institute for Studies in Theoretical Physics and Mathematics (IPM)\\
P.O. Box 19395-5531, Tehran, Iran\\
\vspace*{1mm}
$^b$ Department of Physics, School of Sciences \\
Tarbiat Modares University, P.O. Box 14155-4838, Tehran, Iran}\\

\vspace*{1cm}
\end{center}

\begin{abstract}
We study ${\cal N}=2$ $SO(2N+1)$ SYM theory in the context of matrix model.
By adding a superpotential of the scalar multiplet, $W(\Phi)$, of
degree $2N+2$, we reduce the theory to  ${\cal N}=1$.
The $2N+1$ distinct critical points of $W(\Phi)$
allow us to choose a vacuum in such a way to break the gauge group to
its maximal abelian subgroup.
We compute the free energy of the corresponding matrix model in the
planar limit and up to two vertices. This result is then used to work out
the effective superpotential of ${\cal N}=1$ theory up to one-instanton
correction. At the final step, by scaling the superpotential to zero,
the effective $U(1)$ couplings and the prepotential of the ${\cal N}=2$
theory are calculated which agree with the previous results.

\end{abstract}

\end{titlepage}

\section{Introduction}

The study of ${\cal N}=1$ supersymmetric gauge theories has proven
important in understanding the more realistic theories such as
QCD. This is because on one hand they share many common properties
like chiral symmetry breaking, the existence of a mass gap, and
color confinement in the infrared. And on the other hand,
supersymmetry puts strict, though tractable, conditions on the
dynamics of the theory which makes the theory easier to analyze.
Therefore, a thorough understanding of supersymmetric gauge
theories will help in unraveling the low energy phenomena, of the
kind mentioned above, of the corresponding nonsupersymmetric
theories. This is one, among many others, main reason that
supersymmetric gauge theories are so appealing to study.

A remarkable advance in the understanding of supersymmetric gauge theories
and their relations to Matrix models has recently been achieved through the
work of Dijkgraaf and Vafa \cite{VAFA, MAT, VAFA3, VAFA4}. Consider ${\cal N} =2$
supersymmetric Lagrangian which consists of an ${\cal N} =2$ vector
multiplet $(A , \Phi)$ in the adjoint representation of the gauge group $%
U(N) $. Here $A$ and $\Phi$ are ${\cal N} =1$ vector and chiral multiplets
respectively. Upon adding a superpotential $W(\Phi)$ to the ${\cal N} =2$
Lagrangian, the supersymmetry gets reduced to ${\cal N} =1$. Dijkgraaf and
Vafa have put forward the proposal that the low energy dynamics of this $%
{\cal N} =1$ theory can be completely determined by perturbative
calculations of the free energy of a zero dimensional matrix model in the
planar limit. The potential of the matrix model is taken to be the same as $%
W(\Phi)$, but with $\Phi$'s regarded as constant $M\times M$ matrices in the
Lie algebra of $U(M)$. The most important feature of this correspondence is
that by perturbative calculations in the matrix model side one learns about
the nonperturbative effects -- mainly due to instantons -- in the gauge
theory side. Specifically, let $W(\Phi)$ be a polynomial of degree $n+1$ in $%
\Phi$. The classical supersymmetric vacuum is then characterized by a
constant diagonal matrix with elements $e_i$, the critical points of $%
W(\Phi) $. Let $N_i$ indicate the multiplicity of $e_i$ in the vacuum such
that $N= \sum_i^n N_i$. This choice of vacuum breaks the gauge symmetry as
follows
\[
U(N) \to U(N_1)\times U(N_2)\times\cdots \times U(N_n)\, .
\]
The instantons contributions to the effective superpotential are then given
by
\[
W_{eff}^{{\rm inst}} = -\sum_{i} N_i \frac{\pl {\cal F}_0}{\pl S_i} \, ,
\]
where ${\cal F}_0$ is the free energy of the matrix model in the planar
limit, and $S_i = g_s M_i$.

Using the perturbative calculations in the matrix models, the effective
superpotential of a wide class of ${\cal N} =1$ supersymmetric gauge
theories has been obtained in complete agreement with the earlier results.
Interestingly, one can go even one step further to extract information about
the low energy dynamics of the ${\cal N} =2$ theory itself. This can be done
as follows. One introduces a superpotential $\al W(\Phi)$ of degree $N+1$,
with $\al$ a real parameter, breaking the ${\cal N} =2$ supersymmetry down
to ${\cal N} =1$. Since $W(\Phi)$ has $N$ critical points, one can choose
the vacuum as
\[
\Phi_0 ={\rm diag}\, (e_1 , e_2, \ldots , e_N) \, ,
\]
therefore the $U(N)$ gauge group classically breaks to $U(1)^N$. By adding
the superpotential $W(\Phi)$, one in fact freezes the whole classical vacuum
manifold $C^N$ of ${\cal N} =2$ theory to a point $\Phi_0$, the vacuum of $%
{\cal N} =1$ theory. In conclusion, one computes the effective
superpotential of this theory and notices that there are some low energy
quantities which are independent of the parameter $\al$, and hence must be
identified with the corresponding quantities in the ${\cal N} =2$ theory. In
this way, using the perturbative analysis of the matrix model, the
prepotential of ${\cal N}=2$ $U(2)$ theory was rederived in \cite{KAZA}.
This method was further generalized for the gauge group $U(N)$, and again
with complete agreement with the Seiberg-Witten solution of ${\cal N} =2$ $%
U(N)$ gauge theory \cite{SCH}. It is our aim in this paper to work out the
Seiberg-Witten solution of ${\cal N} =2$ $SO(N)$ gauge theory by
perturbative computations of the free energy of the corresponding matrix
model.

In the above context of gauge theory/matrix model correspondence, ${\cal N}
=1$ $SO/SP$ gauge theories have also been examined from different points of
views \cite{OZ, HAL, FEN, FUJ, OBERS, AHN1, AHN2}. The perturbative matrix 
model language,
though, has only been used to analyze the gauge theory in the trivial vacuum
sector. To derive the ${\cal N} =2$ results, as mentioned above, we
need to choose a vacuum which breaks $SO(2N)$ or $SO(2N+1)$ gauge group to $%
U(1)^N$ representing a typical point on the Coulomb branch of ${\cal N} =2$
vacuum moduli space, and then performing the perturbative calculations
around the corresponding matrix model vacuum. This is what we will do in the
next section.

The organization of this paper is as follows. In section 2, we introduce the
matrix model action including the fluctuations around the vacuum and their ghosts
counterparts which are necessary for our special gauge fixing. In section 3,
we calculate the free energy of the matrix model which consists of three parts; 1)
the nonperturbative part including the contribution of the group volume and
quadratic integrals, 2) two loop planar free energy and 3) unoriented planar
graphs. In section 4, we derive the effective action, and show that the coupling
constants $\tau_{ij}$ can be obtained from the free energy of the matrix model
by a variant form of the Vafa-Dijkgraaf prescription.
The result is then reexpressed in
terms of the periods $a_i$'s. We conclude in section 5 and derive the
Vandermonde determinant of the Fadeev-Popov ghosts in the appendix.


\section{The Matrix Model Superpotential}

In this section, we introduce a superpotential $W(\Phi )$ of the scalar
multiplet which is of degree $2N+2$ for the group $SO(2N+1)$. $W(\Phi )$ is
chosen such that it has $2N+1$ distinct critical points $e_{i}$'s. To be
explicit, let us introduce the superpotential as
\begin{equation}
W(\Phi )=\al \sum_{l=0}^{N}\frac{s_{N-l}(e^{2})}{2l+2}\tr \Phi ^{2l+2}\,,
\label{SUP}
\end{equation}
where
\begin{equation}
s_{m}(e^2)=\sum_{i_{1}<i_{2}<\cdots <i_{m}}e_{i_{1}}^{2}e_{i_{2}}^{2}\ldots
e_{i_{m}}^{2}\,,
\end{equation}
and $\al$ is a real parameter which, at the end, is scaled to zero to read
off the effective $U(1)$ gauge couplings of the ${\cal N}=2$ effective
theory. $e_{i}$'s are the critical points of $W(x)$
\begin{equation}
W^{\prime }(x)=\al x\sum_{l=0}^{N}s_{N-l}(e^{2})x^{2l}=\al %
x\prod_{i=1}^{N}(x^{2}+e_{i}^{2})\equiv \alpha w^{\prime }(x)\,.
\label{WW}
\end{equation}
Taking the vacuum as
\begin{equation}
\Phi _{0}={\mbox {diag}}(0,\ e_{1}i\si_{2},\ e_{2}i\si_{2},\ldots ,\ e_{N}i%
\si_{2})\,,
\end{equation}
will break the gauge group classically as
\begin{equation}
SO(2N+1)\to U(1)^{N}\,. \label{CLASS}
\end{equation}

In the matrix model side, as mentioned before, one takes the same $W(\Phi )$
playing the role of the potential of the model, but with $\Phi $'s now
considered as constant $2M\times 2M$ matrices in the Lie algebra of $SO(2M)$.\footnote{Note that the size of the matrices $M_i$ in the matrix model are chosen
to be large by multiplying $N_i$ by a large number $k_i$. For convenience, we
have taken $k_i$'s to be even so that the matrix model we obtain is even dimensional.} 
To set up the perturbation theory, let us expand the
superpotential around the critical points of $W(\Phi )$. This we do by
substituting $\Phi \to \Phi _{0}+\Psi $ in $W(\Phi )$, where $\Phi _{0}$ is
the vacuum
\begin{equation}
\Phi _{0}={\mbox {diag}}({\bf 0}_{2M_{0}\times 2M_{0}},\ e_{1}i\si%
_{2}\otimes {\bf 1}_{M_{1}\times M_{1}},\ e_{2}i\si_{2}\otimes {\bf 1}%
_{M_{2}\times M_{2}},\ \ldots ,\ e_{N}i\si_{2}\otimes {\bf 1}_{M_{N}\times
M_{N}})\,.  \label{VAC}
\end{equation}
This choice of vacuum will break the gauge group of the matrix model as
follows
\begin{equation}
SO(2M)\to SO(2M_{0})\times U(M_{1})\times U(M_{2})\times \cdots \times
U(M_{N})\,,  \label{BREAK}
\end{equation}
so that
\begin{equation}
M=\sum_{i=0}^{N}M_{i}\,.
\end{equation}
Upon considering the small fluctuations around the vacuum (\ref{VAC}), up to
the second order in $\Psi $, $W(\Phi )$ reads
\begin{equation}
W(\Phi )=\sum_{i=0}^{N}M_{i}W(\imath e_{i})+\frac{1}{2}\alpha
\sum_{l=0}^{N}s_{N-l}\sum_{m=0}^{l}{\rm Tr}(\Psi \Phi _{0}^{m}\Psi \Phi
_{0}^{2l-m})+{\cal O}(\Psi ^{3})\,.
\end{equation}
Further, it is easy to show that the quadratic part is as follows,
\begin{eqnarray}
W_{2} &=&\frac{1}{2}\alpha s_{N}{\rm Tr}(\psi _{00}\psi _{00})+\frac{1}{2}%
\alpha \sum_{i=1}^{N}\sum_{l=0}^{N}ls_{N-l}(\imath e_{i})^{2l}2{\rm Tr}(\psi
_{ii}^{0}\psi _{ii}^{0}-\psi _{ii}^{2}\psi _{ii}^{2})\,,  \nonumber \\
&&
\end{eqnarray}
where we have decomposed the $2M_{i}\times 2M_{j}$ $\ps_{ij}$ matrices in
terms of $\si_{\mu }\,,\mu =0,1,2,3$ matrices, with $\si_{i}$ the Pauli
matrices and $\si_{0}\equiv {\bf 1}_{2\times 2}$,
\begin{equation}
\psi _{ij}\equiv \psi _{ij}^{0}\otimes \sigma _{0}+\psi _{ij}^{1}\otimes
\sigma _{1}+\imath \psi _{ij}^{2}\otimes \sigma _{2}+\psi _{ij}^{3}\otimes
\sigma _{3}\,,
\end{equation}
$\ps_{ij}^{\mu }$ are now $M_{i}\times M_{j}$ matrices.

The important point to notice here is that there are elements of $\Psi $
which are absent in the quadratic part of the action. These include $\psi
_{ij}$ for $i\neq j$ and $\psi _{ii}^{\beta }$ for $\beta =1,3$. Therefore,
these are not propagating fields and one should gauge them away. Note that
the total number of degrees of freedom that we are going to gauge away is
exactly equal to the number of broken gauge generators in (\ref{BREAK}),
i.e.,
\begin{equation}
4M_{0}\sum_{i=1}^{N}M_{i}+4\sum_{i<j}^{N}M_{i}M_{j}+%
\sum_{i=1}^{N}M_{i}(M_{i}-1)\,.
\end{equation}
As we will show in the appendix, the gauge fixing can be implemented by
introducing the Faddeev-Popov ghosts. The ghost action takes the following
form
\begin{equation}
\frac{1}{4}{\rm Tr}B[\Phi ,C]=\frac{1}{4}{\rm Tr}B[\Phi _{0},C]+\frac{1}{4}%
{\rm Tr}B[\Psi ,C]\,.  \label{ghost}
\end{equation}
The kinetic part of the ghost action can be obtained by expanding the ghost
action around the vacuum
\begin{eqnarray}
\frac{1}{4}{\rm Tr}B[\Phi _{0},C] &=&-\sum_{i}{\rm Tr}\left[
(B_{i0}^{1}C_{0i}^{3}-B_{i0}^{3}C_{0i}^{1})-(B_{i0}^{0}C_{0i}^{2}+
B_{i0}^{2}C_{0i}^{0})\right] e_{i} \nonumber \\
&&-\sum_{i}2{\rm Tr}\left[
B_{ii}^{1}C_{ii}^{3}-B_{ii}^{3}C_{ii}^{1}\right] e_{i}
-\sum_{i<j}{\rm Tr}\left[
B_{ji}^{1}C_{ij}^{3}-B_{ji}^{3}C_{ij}^{1}\right](e_{i}+e_{j})\nn \\
&&-\sum_{i<j}{\rm Tr}\left[
B_{ji}^{0}C_{ij}^{2}+B_{ji}^{2}C_{ij}^{0}\right](e_{i}-e_{j}) \,.
\end{eqnarray}

Let us then fix the gauge to $\psi _{ij}=0$ for $i\neq j$, and $\psi
_{ii}^{\beta }=0$ for $\beta =1,3$. Doing so, the interacting part of the
ghost action becomes,
\begin{eqnarray}
\frac{1}{4}{\rm Tr}B[\Psi ,C] &=&\sum_{i<j}{\rm Tr}\left[ (B_{ji}^{\mu }\psi
_{ii}^{0}C_{ij}^{\mu }-B_{ji}^{\mu }C_{ij}^{\mu }\psi _{jj}^{0})\right.
\nonumber \\
&&-(B_{ji}^{1}\psi _{ii}^{2}C_{ij}^{3}+B_{ji}^{1}C_{ij}^{3}\psi
_{jj}^{2})+(B_{ji}^{3}\psi _{ii}^{2}C_{ij}^{1}+B_{ji}^{3}C_{ij}^{1}\psi
_{jj}^{2})  \nonumber \\
&&\left. -(B_{ji}^{2}\psi _{ii}^{2}C_{ij}^{0}-B_{ji}^{2}C_{ij}^{0}\psi
_{jj}^{2})-(B_{ji}^{0}\psi _{ii}^{2}C_{ij}^{2}-B_{ji}^{0}C_{ij}^{2}\psi
_{jj}^{2})\right]  \nonumber \\
&&+2\sum_{i}{\rm Tr}\left[ (B_{ii}^{3}\psi
_{ii}^{2}C_{ii}^{1}-B_{ii}^{1}\psi _{ii}^{2}C_{ii}^{3})+(B_{ii}^{1}\psi
_{ii}^{0}C_{ii}^{1}+B_{ii}^{3}\psi _{ii}^{0}C_{ii}^{3})\right]  \nonumber \\
&&+\sum_{i}{\rm Tr}\left[ (B_{i0}^{3}C_{0i}^{1}\psi
_{ii}^{2}-B_{i0}^{1}C_{0i}^{3}\psi _{ii}^{2})+(B_{i0}^{2}C_{0i}^{0}\psi
_{ii}^{2}+B_{i0}^{0}C_{0i}^{2}\psi _{ii}^{2})\right]  \nonumber \\
&&-\sum_{i}{\rm Tr}(B_{i0}^{\mu }C_{0i}^{\mu }\psi _{ii}^{0})+\frac{1}{2}%
\sum_{i}{\rm Tr}(B_{i0}\psi _{00}C_{0i})\,.
\end{eqnarray}
The kinetic and interaction parts of the ``bosonic'' action, on the other
hand, are found in this gauge to be
\begin{eqnarray}
W_{2} &=&\frac{1}{2}\alpha s_{N}{\rm Tr}(\psi _{00}\psi _{00})+\frac{1}{2}%
\alpha \sum_{i=1}^{N}\sum_{l=0}^{N}(\imath e_{i})^{2l}ls_{N-l}2{\rm Tr}(\psi
_{ii}^{0}\psi _{ii}^{0}-\psi _{ii}^{2}\psi _{ii}^{2}) \\
W_{3} &=&-\alpha \sum_{i=1}^{N}\sum_{l=0}^{N}2l(2l+1)(\imath
e_{i})^{2l-1}(-\imath )s_{N-l}{\rm Tr}(\psi _{ii}^{222}-3\psi _{ii}^{200}) \\
W_{4} &=&-\alpha \frac{s_{N-1}}{4}{\rm Tr}(\psi _{00})^{4}-\alpha
\sum_{i=1}^{N}\sum_{l=0}^{N}\frac{1}{6}(2l-1)2l(2l+1)(\imath
e_{i})^{2l-2}s_{N-l}  \nonumber \\
&&\times 2{\rm Tr}(\psi _{ii}^{2222}-4\psi _{ii}^{2200}-2\psi
_{ii}^{2020}+\psi _{ii}^{0000})\,,
\end{eqnarray}
where we have used the notation $\psi ^{ab\ldots c}=\psi _{{}}^{a}\psi
_{{}}^{b}\ldots \psi _{{}}^{c}$ for $a,b,\ldots =0,2$. Performing the sum
over $l$ we obtain
\begin{eqnarray}
W_{2} &=&\frac{1}{2}\alpha \Delta _{0}{\rm Tr}(\psi _{00}\psi _{00})-\frac{1%
}{2}\alpha \sum_{i=1}^{N}\Delta _{i}{\rm Tr}(\psi _{ii}^{0}\psi
_{ii}^{0}-\psi _{ii}^{2}\psi _{ii}^{2})\,, \\
W_{3} &=&- \imath \alpha \sum_{i=1}^{N}\frac{\gamma _{3,i}}{3}{\rm Tr}(\psi
_{ii}^{222}-3\psi _{ii}^{200})\,, \\
W_{4} &=& -\alpha \frac{s_{N-1}}{4}{\rm Tr}(\psi _{00})^{4}+\alpha
\sum_{i=1}^{N}\frac{\gamma _{4,i}}{4}{\rm Tr}(\psi _{ii}^{2222}-4\psi
_{ii}^{2200}-2\psi _{ii}^{2020}+\psi _{ii}^{0000})\,,
\end{eqnarray}
where use has been made of
\begin{equation}
\sum_{l=0}^{N}ls_{N-l}(\imath e_{i})^{2l}=\frac{\imath e_{i}}{2}\left[ \frac{%
d}{dx}\prod_{k=1}^{N}(x^{2}+e_{k}^{2})\right] _{x=\imath
e_{i}}=-e_{i}^{2}\prod_{k\neq i}(e_{k}^{2}-e_{i}^{2})\,,
\end{equation}
together with the following definitions
\begin{eqnarray}
\Delta _{i} &\equiv &2e_{i}^{2}R_{i},\;\;\;\;\;\;\Delta _{0}\equiv s_{N} \\
\gamma _{3,i} &\equiv &\imath e_{i}R_{i}\left( 3+4e_{i}^{2}\sum_{j\neq i}\frac{1}{%
e_{ij}}\right) \\
\gamma _{4,i} &\equiv &R_{i}\left( 1+8e_{i}^{2}\sum_{j\neq i}\frac{1}{e_{ij}%
}+4e_{i}^{4}\sum_{m\neq i}\sum_{n\neq i,m}\frac{1}{e_{im}e_{in}}\right) \\
R_{i} &\equiv &\prod_{k\neq i}(e_{k}^{2}-e_{i}^{2}),\;\;\;\;\;e_{ij}\equiv
e_{i}^{2}-e_{j}^{2}\,.
\end{eqnarray}
The higher interaction vertices are given by
\be
\ga_{p,i}=\f{1}{(p-1)!} \lf[\lf(\f{\pl}{\pl x}\ri)^{p-1}
x\prod_{j=1}^{N}(x^2+e_j^2)\ri]_{x=\imath e_i}\, .
\ee

With the matrix model perturbative action in hand, now we can find the free
energy of the matrix model by which the gauge theory effective action and
other related quantities are found in the following sections.

\section{Matrix Model Free Energy}

In this section we calculate the free energy ${\cal F}_0$ of the matrix
model which consists of two parts; \newline
1) The non-perturbative part which comes from the volume of the gauge group
and the integration over the quadratic part of the action. \newline
2) The perturbative parts which are coming from the interacting parts of the
matrix model.
Having obtained the free energy ${\cal F}_{0}$,
we use the prescription given by Dijkgraaf and Vafa \cite{VAFA} to write down
the ${\cal N}=1$ effective superpotential $W_{eff}(S)$, which is
\begin{equation}
W_{eff}(S)=-\sum_{i=1}^{N}\left( \frac{\pl {\cal F}_{0}}{\pl S_{i}}+
\la {\cal G}_{0}-2\pi i\tau S_{i}\right) \,,  \label{EFF1}
\end{equation}
where ${\cal G}_0$ is the contributions of the unoriented planar graphs to
the free energy, and $\la =4$ for the $SO(N)$ group \cite{OZ}. $\tau$ is
the bare coupling, and we have set $N_{i}=1$.
Moreover, as in the case of $U(N)$, the effective $U(1)$
couplings can in principle be calculated through the formula
\begin{equation}
2\pi \imath\tau _{ij}(e)=\left( \frac{\partial ^{2}{\cal F}_{0}}{\partial
S_{i}\partial S_{j}}\right) _{{\lan S_{i}\ran}}\,,  \label{TAU}
\end{equation}
where $\lan S_{i}\ran$ are the vev of the gluinos obtained by extremizing
the effective superpotential $W_{eff}(S)$.
However, the case of $SO(N)$ group is a bit subtle and  formula (\ref{TAU})
needs modification.\footnote{This is also noticed in \cite{SCH2} for the
case of $U(N)$ group with matter in (anti)symmetric representation.}
The reason for this is as follows. In the double line notation of t'Hooft,
two index lines of the antisymmetric representations of $SO(N)$ group have
the same orientation (as opposed to the case of adjoint representation of
$U(N)$ group). On the field theory side, since gauge fields $W^\al$ act
through the commutator on matter adjoint fields -- antisymmetric
representations of $SO(N)$ -- one gets an extra minus sign when one moves
one of the $W^\al$ to the outer index loop. These are the
 graphs contributing to the effective $U(1)$ couplings. In order to take
 into account this extra minus sign, in each loop diagram of (anti)symmetric
 field we assign an $S_i$ to one index loop and an $\St_j$ to the adjacent
 index loop. All this amounts to modifying (\ref{TAU}) to
\be
2\pi\imath \tau_{ij}=\lf(\f{\pl^2{\cal F}_0(S , \St)}{\pl S_i \pl S_j}
\right) _{{\lan S_{i}\ran}}\label{TAUU2} \, ,
\ee
noticing that
\be
\f{\pl\St_i}{\pl S_j} =-\del_{ij}\, .\label{RULE}
\ee
And after differentiation setting $\St_i = S_i$.

Knowing the effective $U(1)$ couplings we can proceed to calculate the
prepotential of ${\cal N}=2$ theory. Recall that the ${\cal N}=2$
prepotential is expressed in terms of the periods $a_{i}$'s. Therefore, if
we reexpress (\ref{TAUU2}) in terms of $a_{i}$'s, we can work out the
${\cal N}=2$ prepotential ${\cal F}(a)$ by a double integration of the
following formula
\begin{equation}
\tau _{ij}(a)=\frac{\partial ^{2}{\cal F}(a)}{\partial a_{i}\partial a_{j}}.
\label{prep}
\end{equation}

\subsection{Nonperturbative Part of the Free Energy}

The nonperturbative part of the matrix model free energy ${\cal F}%
_{0}^{({\rm np})} $comprises of three parts. These include the integral over
the kinetic terms of $\psi _{ii}$'s, those of ghosts $B_{ji},C_{ij}$, and the
volume factor of the broken gauge group. Let us discuss each part separately
with some detail. First, the kinetic terms of $\psi $'s consist of three
parts:
\begin{equation}
W_{{\rm {kin}}}(\psi )=\frac{\alpha }{2}\left( -\Delta _{0}{\rm {Tr}(\psi
_{00})^{2}-\sum_{i=1}^{N}\Delta _{i}{Tr}(\psi
_{ii}^{0})^{2}+\sum_{i=1}^{N}\Delta _{i}{Tr}(\psi _{ii}^{2})^{2}}\right) .
\end{equation}
Accordingly, the Gaussian integral over $\psi $'s can be performed easily,
giving the result
\begin{eqnarray}
\int d\psi {\exp \left( -\frac{1}{g_{s}}W_{{kin}}(\psi )\right) } &=&%
{ \left( \frac{2\pi g_{s}}{2\alpha \Delta _{0}}\right)^{\frac{1}{2}
M_{0}(2M_{0}-1)}\prod_i \lf\{
\left( \frac{2\pi g_{s}}{2\alpha
\Delta _{i}}\right)^{\frac{1}{4}M_{i}(M_{i}-1)} \ri. } \nonumber \\
&&\lf. \times
\left( \frac{2\pi g_{s}}{2\alpha \Delta _{i}}\right)^{\frac{1}{4}
M_{i}(M_{i}-1)}\left( \frac{2\pi g_{s}}{\alpha \Delta _{i}}\right)^{\frac{1
}{2}M_{i}}\ri\}.
\end{eqnarray}
Taking into account the appropriate $g_{s}$ factors, and ignoring the linear
terms in $M_{0},M_{i}$ in the planar limit, gives rise to a contribution to $%
{\cal F}_{0}^{({\rm np})}(S)$ of the form

\begin{equation}
S_{0}^2\log \left( \frac{\pi g_{s}}{\alpha \Delta _{0}}\right) +\frac{1}{2}%
\sum_{i}S_{i}^{2}\log \left( \frac{\pi g_{s}}{\alpha \Delta _{i}}\right) .
\end{equation}

Now, we consider the ghost sector. There are three types of ghosts $B,C$
which correspond to the blocks $(ii),(i0,0i),(ij,ji)$ of the original matrix
$\Phi $. As explained in the Appendix, in the eigenvalue representation of
the partition function, the integral over all types of these ghosts produces
the correct Jacobian of the matrix model in the symmetry broken phase,
\begin{equation}
\Delta (\lambda )=\prod_{i}\prod_{\alpha <\beta }(\lambda _{\alpha
}^{(i)}+\lambda _{\beta }^{(i)})^{2}\prod_{i}\prod_{\alpha ,\beta }\left(
(\lambda _{\alpha }^{(0)})^{2}-(\lambda _{\beta }^{(i)})^{2}\right)
^{2}\prod_{i<j}\prod_{\alpha ,\beta }\left( (\lambda _{\alpha
}^{(i)})^{2}-(\lambda _{\beta }^{(j)})^{2}\right) ^{2},
\end{equation}
where $\lambda _{\alpha }^{(i)}$ stands for the eigenvalues in the $i$-th
block. Integrating the kinetic terms of the ghosts then amounts to replacing
the vacuum values $\lambda _{\alpha }^{(i)}=e_{i}, \lambda _{\alpha }^{(0)}=0$
in the above expression.
This will give
\begin{equation}
\int dBdC{\rm e}^{I_{{\rm {kin}}}(B,C;e)}=\prod_{i}(2e_{i})^{M_{i}(M_{i}-1)}%
\prod_{i}(e_{i})^{4M_{0}M_{i}}\prod_{i<j}(e_{ij})^{2M_{i}M_{j}}.
\end{equation}
After inserting the $g_{s}$ factors and ignoring the linear terms in $%
M_{0},M_{i}$, the ghost contribution to ${\cal F}_{0}^{({\rm np})}$ becomes
\begin{equation}
\sum_{i}S_{i}^{2}\log (2e_{i})+4S_{0}\sum_{i}S_{i}\log
e_{i}+2\sum_{i<j}S_{i}S_{j}\log (e_{ij}).
\end{equation}

Finally, let us turn to the volume factor $($vol $G)^{-1}$ for the broken
(matrix model) gauge group $G=SO(2M_{0})\times U(M_{1})\times \cdot \cdot
\cdot \times U(M_{N})$. Using the asymptotic expansion of the volumes of the
groups $SO(2N)$ and $U(N)$ in the large $N$ limit (see
\cite{Ooguri:2002gx, HAL}), we can write it as
\begin{eqnarray}
\log ({\rm {vol}}\, G) &=& \log ({\rm {vol}}\, SO(2M_{0}))+\sum_{i}\log
({\rm {vol}} \, U(M_{i}))  \nonumber \\
&=&-M_{0}^{2}\log M_{0}+\left( \frac{3}{2}+\log \pi \right) M_{0}^{2}+{\cal O%
}(M_{0}\log M_{0})  \nonumber \\
&&+\sum_{i}\left[ -\frac{1}{2}M_{i}^{2}\log M_{i}+\left( \frac{3}{4}+\frac{1%
}{2}\log 2\pi \right) M_{i}^{2}+{\cal O}(\log M_{i})\right] .
\end{eqnarray}
We have kept the next to leading order terms in the above expansion as they
are crucial in cancellation of some numerical factors appearing later. The
contribution of the volume factor to ${\cal F}_{0}^{({\rm np})}$ thus becomes
\begin{equation}
M_{0}^2\log M_{0}+\frac{1}{2}\sum_{i}M_{i}^{2}\log M_{i}-\left( \frac{3}{2}%
+\log \pi \right) M_{0}^{2}-\left( \frac{3}{4}+\frac{1}{2}\log 2\pi \right)
\sum_{i}M_{i}^{2}.
\end{equation}

Summing the above three contributions and the linear terms $%
-\sum_{i}S_{i}W(e_{i})$ coming from the vacuum value of $W(\Phi ),$ we get
the final result for the non-perturbative part of the free energy
\begin{eqnarray}
{\cal F}_{0}^{({\rm np})}(S) &=&-\sum_{i}S_{i}W(e_{i})+S_{0}^{2}\log \left(
\frac{S_{0}}{\alpha \hat{\Lambda}\hat{\Delta}_{0}}\right) +\frac{1}{2}
\sum_{i}S_{i}^{2}\log \left( \frac{S_{i}}{\alpha \hat{\Lambda}^{2}\hat{\Delta
}_{i}}\right)  \nonumber \\
&&+2S_{0}\sum_{i}S_{i}\log \left( \frac{e_{i}^{2}}{\hat{\Lambda}}\right)
+2\sum_{i<j}S_{i}S_{j}\log \left( \frac{e_{ij}}{\hat{\Lambda}}\right) ,
\label{ABOVE}
\end{eqnarray}
where $\hat{\Delta}_{0},\hat{\Delta}_{i}$ are defined as follows
\begin{eqnarray}
e^{-3/2}\hat{\Delta}_{0} &\equiv &\Delta _{0}=R_{0},  \nonumber \\
e^{-3/2}\hat{\Delta}_{i} &\equiv &\frac{\Delta _{i}}{2e_{i}^{2}}=R_{i},
\end{eqnarray}
and $\hat{\Lambda}$ is an arbitrary cut-off. Powers of $\hat{\Lambda}$ are
inserted by hand in the above expression in a way to subtract the overall
term $\left( S_{0}+\sum_{i}S_{i}\right) ^{2}\log \hat{\Lambda}$ from ${\cal F%
}_{0}^{({\rm np})}$. This corresponds to a freedom in choosing the scale of
$\Phi $ in the original model. Indeed, by rescaling $\Phi $ as
$\Phi \rightarrow \sqrt{\Lambda }\Phi ,$ the overall measure of the the
$SO(2M)$ matrix model scales as $d\Phi \rightarrow $
$(\sqrt{\Lambda })^{2M^{2}-M}d\Phi $. This
produces a change in the planar free energy as
\begin{equation}
\delta {\cal F}_{0}=g_{s}^{2}(2M^{2}-M)\log \sqrt{\Lambda },
\end{equation}
which in the t' Hooft limit (with $S=S_{0}+\sum_{i}S_{i}$ a finite quantity)
has precisely the same form $S^{2}\log \Lambda $ as we introduced in Eq.
(\ref{ABOVE}).

As stated earlier, to calculate the effective couplings, we have to rewrite
the free energy (\ref{ABOVE}) by replacing $S_i$ into $\St_i$ wherever
(anti)symmetric fields are present,\footnote{These include
$(B)C^{1,3}_{ii}, (B)C^{1,3}_{ij}$, which in the
double line notation have index lines of the same orientation. But
$\ps_{ii}^{0,2}$ or $(B)C^{0,2}_{ij}$ can be combined into matrices which have
index lines of opposite directions.} i.e.,
\begin{eqnarray}
{\cal F}_{0}^{({\rm np})}(S) &=& \sum_{i}\lf\{
-S_{i}W(e_{i})+\frac{1}{2}
S_{i}^{2}\log  \frac{S_{i}}{\alpha \hat{\Lambda}^{2}{\Delta
}_{i}}  +\f{1}{2} S_i\St_i \log \f{ 2e_i^2}{\hat{\Lambda}} +
2S_{0}\St_{i}\log  \frac{e_{i}^{2}}{\hat{\Lambda}}\ri. \nonumber \\
&&+ \lf. \f{1}{2}\sum_{j\neq i}S_{i}S_{j}\log
 \frac{(e_i-e_j)^2}{\hat{\Lambda}} 
+\f{1}{2}\sum_{j\neq i}S_{i}\St_{j}\log \frac{(e_i+e_j)^2}{\hat{\Lambda}}
 \ri\} \nn \\
&&+ S_{0}^{2}\log  \frac{S_{0}}{\alpha \hat{\Lambda}\hat{\Delta}_{0}} \, .
\end{eqnarray}

Notice that because of the symmetry breaking pattern in (\ref{CLASS}),
a nonzero vev for $S_0$ does not make sense, and therefore we will eventually
set it to zero. However, as we will see shortly, keeping $S_0$ will allow us
to work out a simple rule relating the unoriented contributions ${\cal G}_0$
to the derivative of ${\cal F}_0$ with respect to $S_0$.

\subsection{Two Loop Matrix Model}

Having obtained the propagators and the interaction terms up to the forth
order around the vacuum (\ref{VAC}) in section 2, we are now in a position
to do the perturbative calculations of the free energy ${\cal F}$ in the
planar limit and up to two vertices. Consider the two loops Feynmann
diagrams in Figure 1 and those including ghosts in figure 2.

\bigskip

\begin{figure}[ht]
\begin{center}
\epsfysize=50pt
\epsfbox{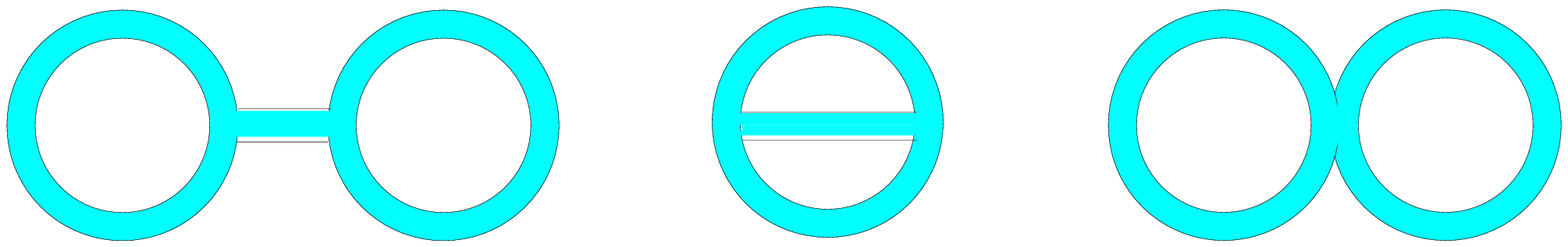}

\vspace{3 mm}
Figure 1) Two loops without ghosts
\end{center}
\end{figure}

\begin{figure}[ht]
\begin{center}
\epsfysize=110pt
\epsfbox{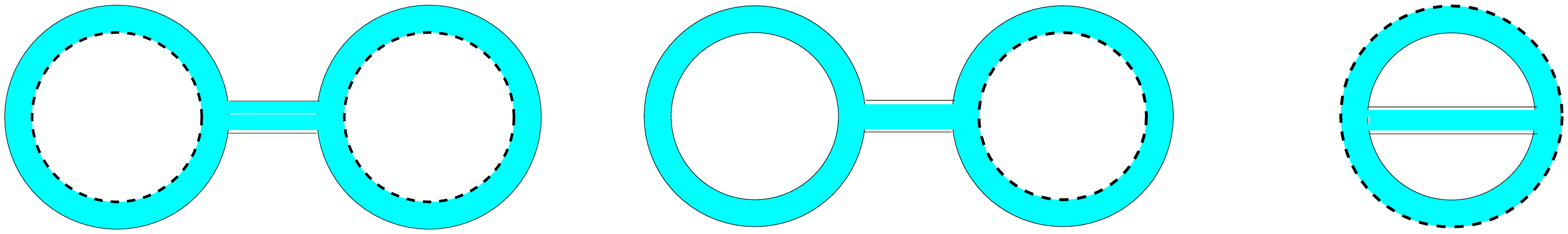}

\vspace{2 mm}
Figure 2) Two loops involving ghosts
\end{center}
\end{figure}
The result of the two loop free energy calculation is
\begin{eqnarray}
{\cal F}_{0}^{(3)} &=&\frac{1}{2}\sum_{i}Y_{i}\left( \frac{1}{2\Delta _{i}}
\right) Y_{i}M_{i}
-\sum_{i}(\frac{1}{6}+\frac{1}{2})\left( \frac{1}{2\Delta _{i}}\right)
^{3}\gamma _{3,i}^{2}M_{i}^{3}  \nonumber \\
&&-2\sum_{i}(\frac{1}{2}+\frac{1}{2}+1)\left( \frac{1}{2\Delta _{i}}\right)
^{2}\gamma _{4,i}M_{i}^{3} -\sum_{i}\left( \frac{1}{2e_{i}}\right) ^{2}\left(
\frac{1}{2\Delta _{i}}\right) M_{i}^{2}\Mt_i  \nonumber \\
&&-\sum_{i}\sum_{j\neq i} \frac{1}{2\Delta _{i}}\lf(
\frac{1}{(e_{i}+e_{j})^{2}}M_i^2\Mt_{j}+
\frac{1}{(e_{i}-e_{j})^{2}}M_i^2M_j\right)
 \nonumber \\
&&-4\sum_{i}\left( \frac{1}{e_{i}}\right) ^{2}\left( \frac{1}{\Delta _{0}}
\right) M_{i}M_{0}^{2}
-2\sum_{i}\left( \frac{1}{e_{i}}\right) ^{2}\left( \frac{1}{2\Delta _i}
\right) M_{i}^2M_{0} \nn \\
&&+ \frac{1}{2}\left( \frac{1}{\Delta _{0}}\right)
s_{N-1}(2M_{0})^{3}  \, ,
\end{eqnarray}
where
\begin{equation}
Y_{i}=\left( \frac{2}{2e_{i}}\right) \Mt_{i}+\left( \frac{2}{2\Delta _{i}}
\right) \imath\gamma _{3,i}M_{i}+\sum_{j\neq i}\lf( \frac{2}{e_{i}-e_j}M_{j}
+\frac{2}{e_{i}+e_j}\Mt_{j}\ri) + \f{4}{e_i}M_0 \, ,
\end{equation}
can be calculated from the tadpole graph.

Restoring the coefficients $\alpha $ and $g_{s}$, and taking $%
S_{i}=g_{s}M_{i} $ we find:
\begin{eqnarray}
\alpha {\cal F}_{0}^{(3)} &=&\sum_{i}\left\{ \lf(-\frac{8}{3}\gamma _{3,i}^{2}
\frac{1}{(2\Delta _{i})^3}-\f{4}{(2\Delta_i)^2}\ga_{4,i}\ri)S_i^3
+\lf( \f{2\imath\ga_{3,i}}{e_i(2\Delta_i)^2}-\f{1}{8e_i^2\Delta_i} \ri) S_i^2\St_i
 \ri. \nn \\
&& +\f{1}{4e_i^2\Delta_i}S_i\St_i^2
-\sum_{j\neq i}\lf( \f{1}{2\Delta_i(e_i+e_j)^2} -\f{4\imath\ga_{3,i}}{(2\Delta_i)^2}
\f{1}{(e_i+e_j)} \ri)S_i^2 \St_j \nn \\
&& -\sum_{j\neq i}\lf( \f{1}{2\Delta_i(e_i-e_j)^2}
-\f{4\imath\ga_{3,i}}{(2\Delta_i)^2} \f{1}{(e_i-e_j)} \ri)S_i^2 S_j  \nn \\
&& + 2\sum_{j\neq i} \lf( \f{1}{2\Delta_i e_i(e_i-e_j)}\St_iS_iS_j
+\f{1}{2\Delta_i e_i(e_i+e_j)}\St_iS_i\St_j \ri) \nn \\
&& + \sum_{j,k \neq i} \lf( \f{2}{2\Delta_i(e_i-e_j)(e_i-e_k)} S_iS_jS_k
+ \f{2}{2\Delta_i(e_i+e_j)(e_i+e_k)} S_i\St_j\St_k \ri) \nn \\
&& +\lf. \sum_{j,k \neq i}\f{4}{2\Delta_i(e_i-e_j)(e_i+e_k)} S_iS_j\St_k \ri\}
+ 4\left( \frac{s_{N-1}}{\Delta _{0}}\right) (S_{0})^{3}\nn \\
&& + \sum_{i}\lf\{ \frac{4}{e_i^2\Delta _{0}}
S_{i}S_{0}^{2} + \f{1}{\Delta_i e_i^2}S_iS_0^2
+\f{\imath\ga_{3,i}}{\Delta_i^2 e_i}S_i^2S_0
+\sum_{j\neq i} \f{4}{\Delta_i (e_i^2-e_j^2)}S_0S_iS_j \ri\} \,. \nn\\
\label{F0}
\end{eqnarray}
Note that $\Delta _{0}\equiv s_{N}=\prod_{i}e_{i}^{2}$ and $\left( \frac{%
s_{N-1}}{s_{N}}\right) =\sum_{i}\frac{1}{e_{i}^{2}}$.

\subsection{Unoriented Planar Contribution to the Free Energy}

Here, we explicitly calculate the unoriented graphs contributions to the
free energy. Notice that since $\ps_{ii}$ (as well as $B_{ii}$ and $C_{ii}$)
matrices are antisymmetric, i.e., take value in the Lie algebra of $SO(M_i)$
, the corresponding propagators must accordingly be antisymmetrized
\begin{equation}  \label{antiprop}
\lan \ps_{\al\bet}\ps_{\ga\del}\ran \sim \frac{1}{2}(\del_{\al\del}\del_{\bet
\ga} -\del_{\bet\del}\del_{\al\ga})\, ,
\end{equation}
where $\al ,\bet ,\ga \ldots =1, \ldots , 2M_i$ indicate the matrix indices.
Therefore, unoriented planar graphs, i.e., graphs with the topology of
sphere with a crosscap, must also be considered in the computation of the
free energy in the planar limit. This will modify the expression for the
effective superpotential to \cite{VAFA, OZ}
\begin{equation}
W_{eff}^{{\rm pert}}(S) = -\sum_{i=0}^N N_i \frac{\pl {\cal F}_0}{\pl S_i} -
\la {\cal G}_0 \, ,
\end{equation}
 Further,
since in the case at hand the gauge group is broken to $U(1)^N$ in the gauge
theory side, we set $N_i =1$ for $i \geq 0$.

We mentioned above that unoriented graphs come from the
anti-symmetrization of the propagators for the antisymmetric matrices (more
precisely from the second term on the r.h.s. of  Eq. (\ref{antiprop})).
In the present
case, due to the decomposition in terms of the Pauli matrices and since $\si
_0 , \si_1 , \si_3$ are symmetric while $\si_2$ is antisymmetric, the
matrices $\psi_{00}$, $\psi_{ii}^0$, $B_{ii}^{1,3}$ and $C_{ii}^{1,3}$
become antisymmetric, whereas $\psi_{ii}^2$ is a symmetric one. Thus the
propagator for $\psi_{ii}^2$ matrices will be,
\begin{equation}
\lan \ps^2_{\al\bet}\ps^2_{\ga\del}\ran \sim \frac{1}{2} (\del_{\al\del}\del
_{\bet\ga} + \del_{\bet\del}\del_{\al\ga})\, .
\end{equation}
As a result it can be seen that the contributions of $\psi^2_{ii}$ and $
\psi^0_{ii}$ to the unoriented part in fact cancel each other. More
interestingly, note that $\psi^2_{ii}$ and $\psi^0_{ii}$ can be put together
to form a hermitian $M_i\times M_i$ matrix
\begin{equation}
\psi_{ii} = \psi^2_{ii} + i\psi^0_{ii}\, .
\end{equation}
This is consistent with the symmetry breaking pattern in (\ref{BREAK}), and
explains why these matrices do not have unoriented graphs.

Next, let us write down the result of the calculations of the unoriented
contributions to the free energy. Effectively, these are coming from a twist
on $\psi _{00}$, $B_{ii}^{1,3}$, and $C_{ii}^{1,3}$ propagators. The
unoriented contribution reads
\begin{eqnarray}
{\cal G}_{0} &=&-\sum_{i}S_{i}\frac{1}{2e_{i}}\frac{1}{\Delta _{i}}%
Y_{i}+\sum_{i}S_{i}^{2}\left( \frac{1}{2e_{i}}\right) ^{2}\frac{1}{\Delta
_{i}}.
\end{eqnarray}
It is easy to show that the above unoriented free energy can be derived by
taking the derivative of the oriented part with respect to $M_{0}$
\cite{OZ, HAL}, i.e.,
\begin{equation}
{\cal G}_{0}=\frac{-1}{2}\frac{\partial }{\partial S_{0}}
{\cal F}_{0}\,.
\label{rule}
\end{equation}
What we have done in this section is a nontrivial illustration of the above
`derivative rule' (\ref{rule}). This rule can be understood naively in some
simpler examples. Putting a twist on a propagator reduces the number of
index loops by one. This has to be done for each loop, and thus, starting
with a graph of order $S^{n}$, we end up with a graph of order $nS^{n-1}$
which is the derivative rule. In our case, however, this naive picture can
not be applied. For example, we see that an unoriented graph can be
constructed by a twist on $B_{ii}C_{ii}$ propagators, while it can be
derived from the derivative of another graph with respect to $S_{0}$. But
our result shows that this comes true!

\section{Effective Superpotential and ${\cal N}=2$ Prepotential from Matrix
Model}

In the previous section, we derived the free energy of the $SO(2M)$ matrix
model in the planar limit and up to two vertices. The prescription given by
Dijkgraaf and Vafa \cite{VAFA} enables us now to write down the ${\cal N}=1$
effective superpotential $W_{eff}(S)$, using (\ref{EFF1}).
In the following subsections, we write down the details of these calculations.

\subsection{Coupling Constants from the Matrix Model}

Let us start by computing $W_{{\rm {eff}}}(S)$ from (\ref{EFF1}) up to order
${\cal O}(S^{3})$. After a little algebra we obtain
\bea \label{EFF2}
W_{{\rm {eff}}}(S)&=& \sum_i \lf(W(e_i) - S_i \log \f{S_i}{\a \hat{\Lambda}^2
\hat{\Delta}_i} - \f{1}{2}S_i + S_i \log \f{4 e_i^2}{\hat{\Lambda}}
- 2 \sum_{j \neq i} S_j \log \f{e_{ij}}{\hat{\Lambda}}  \ri)  \nn\\
&&+\sum_i \lf( \f{-\imath\ga_{3,i}}{\Delta_i^3}-
\f{3 \ga_{4,i}}{\Delta_i^2}+\f{3\imath \ga_{3,i}}{2e_i\Delta_i^2}
+\f{3}{8e_i^2\Delta_i}\ri) S_i^2 \nn\\
&& + \sum_i \sum_{j \neq i} \lf( -\f{2 (e_i^2+e_j^2)}{\Delta_i e_{ij}^2}+
\f{2\imath \ga_{3,i}}{\Delta_i^2}\f{e_i}{e_{ij}}+\f{2}{\Delta_i e_{ij}}\ri)
(2 S_i S_j+S_i^2) \nn\\
&& + \sum_i \sum_{j \neq i}\sum_{k \neq i} \lf( \f{4 e_i^2}
{\Delta_i e_{ij}e_{ik}} \ri)( S_i S_j+ S_i S_k+S_j S_k) + 4 {\cal G}_0  \; .
\eea
Upon extremizing $W_{{\rm {eff}}}(S),$ it is found that
\begin{equation} \label{sequation}
\f{\pl W_{{\rm {eff}}}}{\pl S_m}= \log \left( \frac{S_{m}}{\alpha
\hat{\Lambda}^2 \hat{\Delta}_m}\right) + \f{3}{2}-
\log \f{4e_m^2}{\hat{\Lambda}}
+ 2 \sum_{j \neq m} \log \f{e_{jm}}{\hat{\Lambda}}
+ \frac{1}{\alpha }\sum_{j}A_{mj}S_{j} =0 \; ,
\end{equation}
where $A_{mj}$ denote the coefficients of the quadratic part of
$W_{{\rm {eff}}}(S)$ in  (\ref{EFF2}).

For small $\hat{\Lambda}$, we can solve the equation (\ref{sequation})
by iteration to find the roots $S_{m}=\langle S_{m}\rangle $. The result up to
the second order is given by
\begin{equation}
\langle S_{m}\rangle =\alpha \tilde{\Lambda}\f{4e_m^2}{R_m}-\alpha
\tilde{\Lambda}^{2}\sum_{j}A_{mj}\f{4e_m^2}{R_m}\f{4e_j^2}{R_j},
\end{equation}
where we have defined the new cut-off $\tilde{\Lambda}$ as
\begin{equation}
\tilde{\Lambda}\equiv \hat{\Lambda}^{2(N+n_{0})}\exp \left( 2i\pi \tau
_{0}\right) \equiv \Lambda ^{2(N+n_{0})}\, .
\end{equation}
The real gauge theory cut-off $\Lambda $ is the one defined by the last
equality.

We note that the perturbative and $d$-instanton parts of $\tau _{mn}$ in the
above decomposition come from ${\cal F}_{0}^{({\rm np})}$ and ${\cal F}%
_{0}^{(d+2)} $ terms in the matrix model side, respectively. By
differentiating ${\cal F}_{0}^{({\rm np})}(S)$ and ${\cal F}_{0}^{(3)}(S)$
according to the rules (\ref{TAUU2}, \ref{RULE}), at the point $S_i=\langle
S_{i}\rangle $,
one can find $\tau _{mn}^{({\rm {pert})}}$ and $\tau _{mn}^{(1)}$
in terms of $e_{i}$'s,
\begin{eqnarray} \label{TIJ1}
2\pi \imath\tau _{mn}^{({\rm {pert})}} &=&\delta_{mn}\lf\{
-2\sum_{i}\log\f{e_{im}}{\tilde{\Lambda}} - \log \f{4 e_m^2}{\hat{\Lambda}}
 \ri\} +(1 - \del_{mn})\log \f{(e_n-e_m)^2}{(e_n+e_m)^2} \; ,\\
2\pi \imath\tau _{mn}^{(1)} &=& \delta_{mn}\lf\{\lf(\f{-2\ga_{3,m}^2}{\Delta_m^3}-
\f{6 \ga_{4,m}}{\Delta_m^2}-\f{\imath\ga_{3,m}}{e_m\Delta_m^2}
-\f{1}{4e_m^2\Delta_m}\ri)\f{4e_m^2}{R_m} \ri. \nn\\
&& \lf. +  \sum_{j \neq m} \f{4e_j^2}{R_j}\lf( -\f{2 (e_m^2+e_j^2)}
{\Delta_m e_{mj}^2}+\f{4\imath\ga_{3,m}}{\Delta_m^2}\f{e_m}{e_{mj}}+
\f{8 e_m^2 }{\Delta_j e_{mj}}\ri)-\sum_{j}A_{mj}\f{4e_m^2}{R_m}\f{4e_j^2}{R_j}
 \ri\} \nn\\
&&+ (1-\delta_{mn}) \lf\{ \f{16e_m^2 e_n}{\Delta_mR_m}\lf(
\f{-e_m}{ e_{mn}^2}+ \f{\imath \ga_{3,m}}{\Delta_m e_{mn}}\ri)
+ \f{16 e_n^2 e_m}{\Delta_nR_n}\lf(
\f{- e_n}{ e_{mn}^2}+ \f{\imath \ga_{3,n}}{\Delta_n e_{nm}}\ri)
\ri. \nn\\
&& \lf. +\sum_{j \neq m} \f{8 e_m e_n}{\Delta_m e_{mn}e_{mj}}\f{4e_j^2}{R_j}
+\sum_{j \neq n} \f{8 e_m e_n}{\Delta_n e_{nm}e_{nj}}\f{4e_j^2}{R_j}
+\sum_{j \neq m,n} \f{8 e_m e_n}{\Delta_j e_{jm}e_{jn}}\f{4e_j^2}{R_j}\ri\}
\label{TIJ2}
\end{eqnarray}

As expected, these quantities turn out to be independent of the
parameter $\alpha $. Therefore, the coupling constants of the unbroken $U(1)$
factors of the ${\cal N}=2$ gauge theory are given by the $mn$ components of
the above equation.

\subsection{ Computation of the Periods within the Matrix Model}

In ref.\cite{SCH} a method was proposed for the computation of the
periods $a_{i}$ of the Seiberg-Witten curve. The method is in fact based on
a purely perturbative calculation of the planar tadpole diagrams within the
matrix model with no reference to the actual form of the Seiberg-Witten curve
or differential. Here, within the same framework as in \cite{SCH}, we use a
rather different method based on differentiating with respect to the
variation of the potential of the matrix model by linear source terms. To be
specific, let us consider the original matrix model with linear source terms
of the form $-\sum_{i}\epsilon _{i}$Tr$(\phi _{ii}^{2})$, with $\epsilon
_{i} $ infinitesimal parameters. The planar free energy of this
modified matrix model is given by the following equation
\begin{equation}
\exp \left( \frac{1}{g_{s}^{2}}{\cal F}_{0}^{\prime }\right) =\frac{1}{{\rm {
vol\,}G}}\int d\Phi \exp \left( -\frac{1}{g_{s}}\left( W(\Phi
)-\sum_{i}\epsilon_{i}{\rm Tr}(\phi _{i})\right) \right) _{{\rm {planar}}},
\end{equation}
where we have put\footnote{
Note that we do not need to consider a source for $\phi _{00}$
block, since it is an antisymmetric matrix and has Tr$(\phi _{00})=0,$
corresponding to $a_{0}=0$. Also $\phi_{ii}^0$ has zero trace.}
$\phi _{i}\equiv \phi _{ii}^{2}$. After all,
this implies a simple relation between the planar tadpole diagrams given by $
\langle $Tr$(\phi _{i})\rangle _{0}$ and the free energy as\footnote{For
the precise definition of the operators $\frac{\delta }{\delta \epsilon
_{i}}$ see below.}
\begin{equation}
\langle Tr(\phi _{i})\rangle _{0}=\frac{1}{g_{s}}\frac{\delta {\cal F}_{0}}{%
\delta \epsilon _{i}}.
\end{equation}

Adding the source terms amounts to replacing the block superpotentials by
\begin{equation}
{\rm {Tr}}w(\phi _{i})\rightarrow {\rm Tr}\tilde{w}_{i}(\phi
_{i})\equiv {\rm Tr} w(\phi_{i})-\epsilon_{i}{\rm Tr}(\phi_{i}),
\end{equation}
in which
\bea
W(\phi)&\equiv& \sum_i {\rm Tr} w(\phi_i) \nn\\
\tilde{w}_{i}(x)&\equiv& w(x)-\epsilon _{i}x.
\eea
This modification clearly changes the vacuum of the matrix model. The true
shift in the vacuum can be easily obtained by going to the eigenvalue
representation of the matrix model. In this representation the vacuum values
of $\lambda^{(i)}$'s in different blocks are determined by extremizing the
associated superpotentials, that is for the $ii$ block by the equation
\begin{equation}
w^{\prime }(x)=\epsilon _{i}\, .
\end{equation}
This change in the vacuum causes the zero point energies, the couplings, and
the propagators of the original matrix model shift according to the following
relations\footnote{
The $l=2$ choice in the last line of these equations corresponds to a
modification of the propagators of $\psi _{i}$, while the $l>2$ choice
gives the changes in their vertex factors.}
\begin{eqnarray}
e_{i} &\rightarrow &\tilde{e}_{i}=e_{i}+\frac{\epsilon _{i}}{w^{\prime
\prime }(e_{i})},  \nonumber \\
w(e_{i}) &\rightarrow &\tilde{w}_{i}(\tilde{e}_{i})=w(e_{i})-\epsilon
_{i}e_{i},  \nonumber \\
w^{(l)}(e_{i}) &\rightarrow &\tilde{w}_{i}^{(l)}(\tilde{e}%
_{i})=w^{(l)}(e_{i})+\epsilon _{i}\frac{w^{(l+1)}(e_{i})}{w^{\prime \prime
}(e_{i})},\qquad l\geq 2.
\end{eqnarray}
It is important to note that, although the quantities $w^{(l)}(e_{i})$ are
explicit functions of all $e_{i}$'s, in the above procedure, we
have replaced only $e_{i}$ in the argument of $w^{(l)}(x),$ holding all
its ($e_{i}$-dependent) coefficients fixed. Since the planar free
energy ${\cal F}_{0}$ is in general a function of $S_{I}$ and the parameters
$e_{i},w(e_{i}),w^{(l)}(e_{i})$, the above discussion indicates that
the addition of the source terms has the net effect of
changing ${\cal F}_{0}$ as follows
\begin{equation}
{\cal F}_{0}\left( S_{I},e_{i},w(e_{i}),w^{(l)}(e_{i})\right) \rightarrow
{\cal F}_{0}^{\prime }\equiv {\cal F}_{0}\left( S_{I},\tilde{e}_{i},\tilde{w}%
_{i}(\tilde{e}_{i}),\tilde{w}_{i}^{(l)}(\tilde{e}_{i})\right) .
\end{equation}
In particular, this shows that the differential operator $\frac{\delta }{\delta \epsilon
_{i}}$ must be defined precisely as follows
\begin{equation}
\frac{\delta }{\delta \epsilon _{i}}\equiv -e_{i}\frac{\partial }{\partial
w(e_{i})}+\frac{1}{w^{\prime \prime }(e_{i})}\left( \frac{\partial }{%
\partial e_{i}}+\sum_{l\geq 2}w^{(l+1)}(e_{i})\frac{\partial }{\partial
w^{(l)}(e_{i})}\right) .
\end{equation}

Now, we turn to the calculation of the Seiberg-Witten periods. By the same
line of reasoning as in \cite{SCH}, we
define the periods $a_{i}$ in the matrix model using the following equation (It can also be computed in terms of the planar tadpole diagrams)
\begin{equation}
a_{i}=g_{s}\sum_{K=0}^{N}n_{K}\left( \frac{\partial }{\partial S_{K}}\langle
Tr(\phi _{i})\rangle _{0}\right) _{\langle S\rangle }\, ,\label{TAD}
\end{equation}
where $n_{I}$ are defined as follows:
\bea
&&n_i=1, \;\;\; i=1 \dots N \nn\\
&&n_0=\f{N_0}{2}-1.
\eea
Note that for $SO(2N)$, $n_0=-1$, and for $SO(2N+1)$, $n_0=-1/2$.

Upon expanding around the vacuum, $\phi _{i}=e_{i}+\psi _{i}$,
one sees that $\langle Tr(\phi _{i})\rangle _{0}=e_{i}M_{i}+\langle
Tr(\psi _{i})\rangle$. Eq. (\ref{TAD}) now implies the expected
expansion of $a_{i}$ as $a_{i}=e_{i}+{\cal O}(\langle S\rangle ).$ Using the
relation $\langle Tr(\phi _{i})\rangle _{0}=\frac{1}{g_{s}}\frac{\delta
{\cal F}_{0}}{\delta \epsilon _{i}}$ for the tadpole, we can rewrite $%
a_{i}$ as
\begin{equation}
a_{i}=\left( \frac{\delta }{\delta \epsilon _{i}}\sum_{K=0}^{N}n_{K}\frac{%
\partial {\cal F}_{0}}{\partial S_{K}}\right) _{\langle S\rangle }.
\end{equation}
Noticing the general formula for $W_{{\rm {eff}}}$ in terms of ${\cal F}%
_{0} $, we are led to the final general formula for $a_{i}$%
\begin{equation}
a_{i}=\left( \frac{\delta W_{{\rm {eff}}}}{\delta \epsilon _{i}}\right)
_{\langle S\rangle }=\frac{\delta }{\delta \epsilon _{i}}W_{{\rm {eff}}%
}(\langle S\rangle ),  \label{ai}
\end{equation}
where in the last step, we have used the fact that $\partial W_{%
{\rm {eff}}}/\partial S=0$ at $\langle S\rangle $. In using the above formula,
we should be careful to use the full expression of $W_{{\rm {eff}}%
}(\langle S\rangle )$ in terms of $e_{i}$ and $w^{(l)}(e_{i})$ without
using the explicit forms of $w^{(l)}(e_{i})$ in terms of $e_{i}$'s.

In the case of our interest,
 $W_{{\rm {eff}}}(\langle S\rangle )$ can be expressed as
functions of $e_{i}$, $w^{\prime \prime }(e_{i})$, $w^{(3)}(e_{i})$, and $w^{(4)}(e_{i})$%
. Thus our formula (\ref{ai}) up to the order $\tilde{\Lambda}$ is simplified
to
\begin{equation}
a_{i}=e_{i}-\frac{\tilde{\Lambda}}{w^{\prime \prime }(e_{i})}\left( \frac{%
\partial }{\partial e_{i}}+w^{(3)}(e_{i})\frac{\partial }{\partial w^{\prime
\prime }(e_{i})}\right) W_{\rm eff}^{(1)}+{\cal O}(\tilde{%
\Lambda}^{2}).
\end{equation}
where $W_{\rm eff}^{(1)}$ is the first order term  of $W_{\rm eff}$.
Thus for $a_{i}$ we find
\begin{equation}
a_{i}=e_{i}-\frac{2\tilde{\Lambda}}{w^{\prime \prime }(e_{i})}\left(
(2-4n_{0})\frac{(e_{i})^{1-4n_{0}}}{w^{\prime \prime }(e_{i})}-w^{(3)}(e_{i})%
\frac{(e_{i})^{2-4n_{0}}}{(w^{\prime \prime }(e_{i}))^{2}}\right) +{\cal O}(%
\tilde{\Lambda}^{2}).
\end{equation}

Finally for $SO(2N+1)$ case, we find $e_i$'s in terms of $a_i$'s as follows:
\bea \label{E}
e_{i}=a_{i}+\tilde{\Lambda} f_i
+{\cal O}(\tilde{\Lambda}^{2})\, ,
\eea
where
\bea 
f_i&=&\f{1}{4 a_i R_i^2}\lf( 1-4 a_i^2\sum_{j\neq i}\f{1}{a_{ij}} \ri) \\
R_i&=& \prod_{k\neq i} (a_k-a_i)\; .
\eea
For determining the prepotential ${\cal F}(a)$, we need to express $\tau
_{ij}(e)$ in terms of $a_{i}$ instead of $e_{i}$ using (\ref{E}).
Thus we obtain,
\begin{equation}
\tau _{ij}(a)=\tau _{ij}^{({\rm {pert})}}(a)+\tilde{\Lambda}\left(
\tau_{ij}^{(1)}(a)+ \sum_k\frac{\partial \tau _{ij}^{({\rm pert})}(a)}
{\partial a_{k}}f_{k}(a)\right)
+{\cal O}(\tilde{\Lambda}^{2}).
\end{equation}
Here, $\tau _{ij}^{({\rm {pert})}}(a)$ and $\tau _{ij}^{(1)}(a)$ are given
by replacing $e_{i}\rightarrow a_{i}$ in (\ref{TIJ1}) and (\ref{TIJ2}).
The final results are as follows,
\begin{eqnarray}
2\pi i\tau _{mn}^{({\rm pert})} &=&\delta_{mn}\lf\{
-2\sum_{i}\log\f{a_{im}}{\tilde{\Lambda}} - \log \f{4 a_m^2}
{\hat{\Lambda}} \ri\}
+(1 - \del_{mn})\log \f{(a_n-a_m)^2}{(a_n+a_m)^2} \; ,
\end{eqnarray}
\begin{eqnarray}
2\pi i\tau _{mn}^{(1)} &=& \delta_{mn}\sum_{j\neq m}
\lf( \f{-4}{R_m^2 a_{mj}}+\f{24a_m^2}{R_m^2 a_{mj}^2}
+\f{24a_m^2}{R_j^2 a_{mj}^2}-\f{-4}{R_j^2 a_{mj}}
+\f{16 a_m^2}{R_m^2}\sum_{k\neq m,j}\f{1}{a_{mj} a_{mk}} \ri) \nn\\
&&+(1-\delta_{mn}) \lf\{ \f{-8a_m a_n}{a_{mn}^2}\lf(
\f{1}{R_m^2}+ \f{1}{R_n^2}\ri)
+ 16a_m a_n \sum_{j\neq m,n}\f{1}{R_j^2 a_{nj}a_{mj}} \ri. \nn\\
&& \lf. \hspace{24mm} -\f{16a_m a_n}{a_{nm}}\lf(
\f{1}{R_n^2}\sum_{j\neq n} \f{1}{a_{nj}}-\f{1}{R_m^2}\sum_{j\neq m}
\f{1}{a_{mj}} \ri)\ri\}\, .
\end{eqnarray}

It is easy to show that the above expression for $\tau _{mn}(a)$
can be integrated to give the following prepotential
${\cal F}(a),$ up to the one-instanton correction:
\bea
{\cal F}(a)^{\rm (pert)}&=& \f{\imath}{4 \pi} \lf\{ \sum_l \sum_{k \neq l}
\lf( (a_k+a_l)^2 \log\f{(a_k+a_l)^2}{\hat{\Lambda}}+ (a_k-a_l)^2
\log\f{(a_k-a_l)^2}{\hat{\Lambda}}\ri)\ri. \nn\\
&&\lf.\hspace{10mm}+2\sum_k a_k^2 \log \f{a_k^2}{\hat{\Lambda}} \ri\}, \\
{\cal F}(a)^{(1)}&=& \f{\tilde{\Lambda}}{16 \pi\imath} \sum_k \prod_{l\neq k}
\f{1}{(a_k^2-a_l^2)^2} \; .
\eea
which is in agreement with
the known results in the ${\cal N}=2$ theory \cite{D'Hoker:1996mu}.

\section{Conclusion}

We studied the ${\cal N}=2$ theory with the gauge group $SO$ using the
Dijgkraaf-Vafa proposal of Matrix Model approach to the ${\cal N}=1$ SYM
theories. This was done by adding a superpotential to the ${\cal N}=2$
theory which broke it to ${\cal N}=1$, then using the corresponding matrix
model, we computed the effective action for ${\cal N}=1$ gauge theory, with
a nontrivial vacuum breaking the group into its maximal abelian subgroup. We
chose this vacuum as we were interested in finding the ${\cal N}=2$
prepotential in the Coulomb branch. For this reason, and to derive the $%
{\cal N}=2$ effective couplings, we finally turned off the superpotential by
sending its coefficient $\al$ to zero. As expected, the coupling constants $%
\tau_{ij}$ were independent of $\al$ and thus were identified with the $%
{\cal N}=2$ effective $U(1)$ couplings. At the end, $\tau_{ij}$ were
integrated to find out the prepotential of ${\cal N}=2$ theory.

In the calculation of the effective action, we carefully considered the
unoriented graphs of the anti-symmetric matrices, and observed that their
contributions can be rederived from the derivative of planar graphs with
respect to the supergluball field, $S_0$. This provided an interesting and
nontrivial example for the `derivative rule'.

We also computed the periods of ${\cal N}=2$ theory by adding a source
term to the matrix model action. This is equivalent to computing the tadpole
graphs. However, the calculation of periods we did is general enough to be
used in similar matrix models.

The extension to $SP(2N)$ and $SO(2N)$ gauge groups is straightforward.
For $SO(2N)$ group, the calculation steps are very much similar to that
of $SO(2N+1)$, though, a complication may arise due to the presence of
Pfaff($\phi$) in the superpotential.

{\large {\it Note Added}}.
During the course of this investigation the paper \cite{OOK} appeared which
considers ${\cal N}=1$ $SO/SP$ gauge theories.
They have derived the effective action.

\hspace{-6mm}

{\large {\bf Acknowledgement}} \newline
We would like to thank M. Alishahiha for useful comments and discussions. We
are also grateful to C. Vafa for valuable discussions on his proposal.


\appendix

\section{Appendix}

In this appendix, we will show that the Faddeev-Popov ghost action needed to
fix the gauge to $\Phi_{ij}=0$, $i\neq j$  is
the one given by (\ref{ghost}). To begin with, let us first diagonalize the
matrix $\Phi$ by an orthogonal $SO(2M)$ transformation, and call the
eigenvalues $\la_I$,
\begin{equation}
\Phi = {\mbox {diag}}( \la_1 i\si_2 , \ldots , \la_{M} i\si_2)\, .
\end{equation}
The superpotential (\ref{SUP}) is thus
\begin{equation}
W(\la)=2\al \sum_{l=0}^{N} \sum_{I=1}^{M}\frac{s_{N-l}(e^2)}{2l+2} \la%
_I^{2l+2} \, .  \label{SUPER}
\end{equation}
Further, if we define
\begin{equation}
\ph_i \equiv {\mbox {diag}} (\la_1 (i\si_2), \la_2 (i\si_2),\ldots , \la%
_{M_i}(i\si_2))\, ,  \label{DIA}
\end{equation}
then
\begin{equation}
\Phi ={\mbox {diag}} (\ph_1,\ldots ,\ph_N )\, .
\end{equation}
The superpotential (\ref{SUPER}) can now be written as
\begin{equation}
W(\Phi)=\al \sum_{l=0}^{N}\sum_{I=1}^{N} \frac{s_{N-l}(e^2)}{2l+2}\tr %
\Phi_I^{2l+2} \, .
\end{equation}

In diagonalizing the $\Phi$ matrix, one also has to take into account the
Vandermonde determinant, which appears in the measure as the Jacobian of the
transformation. For the group $SO(2M)$, this determinant reads
\begin{equation}
\Delta = \prod_{I\neq J}^{M} (\la_I^2 -\la_J^2)= \Delta^{(1)}\cdot
\Delta^{(2)}\, ,
\end{equation}
where
\begin{eqnarray}
&& \Delta^{(1)} = \prod_{I_1\neq J_1}^{M_1} (\la_{I_1}^2 -\la_{J_1}^2)
\prod_{I_2\neq J_2}^{M_2} (\la_{I_2}^2 -\la_{J_2}^2) \ldots \prod_{I_{N}\neq
J_{N}}^{M_{N}} (\la_{I_{N}}^2 -\la_{J_{N}}^2)  \nonumber \\
&& \Delta^{(2)}= \prod_{ I_i, J_j (i\neq j)}^{} (\la_{I_i}^2 - \la_{J_j}^2)
\, .
\end{eqnarray}
Let us now write the second part of the Vandermonde determinant $%
\Delta^{(2)} $ as an integral over ghosts. First note that for a fixed $\la%
_1 $ and $\la_2$ we have
\begin{equation}
(\la_{1}^2 - \la_{2}^2)^2 = \int dB_{2 1} dC_{1 2} \exp\left( B_{2 1}^{\al%
\bet}\la_{1}(i\si_2)_{\bet\al}C_{1 2}^{\al\bet} + C_{1 2}^{\al\bet}\la_{2}(i%
\si_2)_{\bet\al} B_{2 1}^{\al\bet} \right)\, ,
\end{equation}
where $\al ,\bet =1, 2$. Therefore
\begin{eqnarray}
\prod_{I_1,J_2}^{M_1, M_2}(\la_{I_1}^2- \la_{J_2}^2)^2 &=& \int
\prod_{I_1,J_2}dB_{J_2 I_1} dC_{I_1 J_2}  \nonumber \\
&\times &\!\!\! \exp\left(\sum_{I_1,J_2}^{M_1,M_2} B_{J_2 I_1}^{\al\bet} \la%
_{I_1}(i\si_2)_{\bet\al} C_{I_1 J_2}^{\al\bet} + C_{I_1 J_2}^{\al\bet}\la%
_{J_2}(i\si_2)_{\bet\al} B_{J_2 I_1}^{\al\bet}\right)\, .  \label{I}
\end{eqnarray}
Using definition (\ref{DIA}), this can be written as
\begin{equation}
\prod_{I_1,J_2}(\la_{I_1}^2 - \la_{J_2}^2)^2 = \int dB_{2 1} dC_{1 2}
\exp\left( \tr_2 (B_{2 1}\ph_{1}C_{12}) + \tr_1(C_{12}\ph_{2}B_{21})\right)%
\, ,
\end{equation}
where the subindex $i$ indicates the trace is over $2M_i\times 2M_i$
matrices. It is also understood that $B_{ji}$ and $C_{ij}$ are $2M_j\times
2M_i$ and $2M_i\times 2M_j$ matrices, respectively. The Vandermonde
determinant $\Delta^{(2)}$ now reads
\begin{equation}
\prod_{I_i, J_j}^{} (\la_{I_i}^2 - \la_{J_j}^2)^2 = \int \prod_{i< j}dB_{ji}
dC_{ij} \exp\left(\sum_{i< j}\tr_j( B_{ji}\ph_{i}C_{ij}) + \tr_i (C_{ij}\ph%
_{j}B_{ji})\right)\, .
\end{equation}
Therefore, the partition function turns out to be
\begin{eqnarray} \label{ACT}
Z = \int d\Phi dB dC \exp\left( \al \sum_{l,i} \frac{
s_{N-l}(e^2)}{2l+2}\tr \ph_i^{2l+2} + \sum_{i< j}\tr_j( B_{ji}\ph
_{i}C_{ij}) + \tr_i (C_{ij}\ph_{j}B_{ji})\right)
\end{eqnarray}
where the measure is
\begin{equation}
d\Phi dB dC = \prod_{I}d\la_I \prod_{I_1\neq J_1}^{M_1} (\la_{I_1}^2 -\la%
_{J_1}^2) \prod_{I_2\neq J_2}^{M_2} (\la_{I_2}^2 -\la_{J_2}^2) \ldots
\prod_{I_N\neq J_N}^{M_N} (\la_{I_N}^2 -\la_{J_N}^2) \prod_{i< j}dB_{ji}
dC_{ij} \, .  \label{MES}
\end{equation}

With the Vandermonde determinant $\Delta^{(1)}$ in the measure (\ref{MES}),
one cannot go very far in perturbation theory. However, $\Delta^{(1)}$ can
be re-absorbed in the action; simply drop the determinant, and in effect
change the $\si_2$-diagonal $\ph_i$ matrices into some $2M_i\times 2M_i$
matrices $\ph_{ii}$ with $\la_{I_i}$'s as their eigenvalues. At the end, the
partition function will be
\begin{equation}
Z = \int \prod_{i}d\ph_{ii} \prod_{i< j}dB_{ji} dC_{ij} \exp\left( W(\Phi) +
\sum_{i< j}^{N} \tr_j( B_{ji}\ph_{ii}C_{ij}) + \tr_i (C_{ij}\ph%
_{jj}B_{ji})\right)\, .
\end{equation}
Noticing that $B_{ji}=-B_{ij}^T$ and $C_{ji}=-C_{ij}^T$ (as $B$ and $C$ are $%
SO(2M)$ Lie algebra valued), the ghost action can be written as
\begin{equation}
S_{{\rm gh}} = \frac{1}{2} B[\Phi , C] \, ,
\end{equation}
which is the same action written in (\ref{ghost}).

\end{document}